\documentstyle[aps,twocolumn]{revtex}

\begin{document}
\date{\today}
\draft

\title{Role of Self-Interaction Effects in the
Geometry Optimization of Small Metal Clusters}

\author{J. M. Pacheco$^{a)}$, W. Ekardt$^{b)}$, and W.-D. Sch\"one$^{b)}$}

\address{
a) Departamento de Fisica da Universidade \\
3000 Coimbra, Portugal \\
b) Fritz-Haber-Institut der Max-Planck-Gesellschaft,\\
Faradayweg 4-6, 14195 Berlin, Germany}

\maketitle

\begin{abstract}
By combining the Self-Interaction Correction (SIC) with
pseudopotential perturbation theory,
the role of self-interaction errors inherent to the Local Density
Approximation (LDA) to Density Functional Theory is
estimated in the determination of ground state and low energy
isomeric structures of small metallic clusters.
Its application to
neutral sodium clusters with 8 and 20 atoms shows
that the SIC provides
sizeable effects in $Na_8$, leading to a different ordering of the low lying
isomeric states compared with {\it ab-initio} LDA predictions, whereas
for $Na_{20}$, the SIC effects are less pronounced, such that
a quantitative agreement is achieved between the present
method and {\it ab-initio} LDA calculations.
\end{abstract}

\pacs{36.40.+d, 31.20.Sy,31.50.+w, 33.20.Kf}


The use of Density Functional Theory (DFT), together with the
Local (Spin) Density Approximation (L(S)DA) to perform
structural optimization of small finite systems as well as to
predict the stability of new, finite and infinite, materials,
has increased dramatically in recent years, with a widespread use in Physics
and Chemistry\cite{jena}. In
this framework, the problem reduces to the
repeated determination of the
self-consistent solution of the Kohn-Sham equations.
Since in many materials (such as metals and the metallic
clusters considered in this work) the interaction between
the valence electrons and the ionic cores is weak - due to
the effective screening of the nuclear charge by the
tightly bound core electrons -
the resulting electron-ion interaction can be formulated in
terms of pseudopotentials. Several
{\it ab-initio} norm-conserving pseudopotentials are available which
ensure a high degree of transferability,
and are obtainable either via analytic formulae, or
their numerical tabulation requires a few CPU seconds\cite{hamann,tm2}.
Indeed, {\it ab-initio}
LSDA-DFT methods, combined with norm conserving pseudopotentials,
have been recently used to perform structural minimization
of clusters with as many as 147 atoms\cite{weare}.
\par
Other {\it ab-initio} methods, such as the Self-Consistent-Field
Configuration Interaction (SCF-CI)\cite{vlasta1}, which typically treat as
active all the electrons of each atom, are more limited in cluster size,
due to their higher computational cost, and also
due to basis-set\cite{perzun,wanda} and size-consistency\cite{kaldor}
problems which preclude, e.g.,
the study of cluster trends as a function of size.
However, these methods have no
self-interaction problems, such as those well-kown in LDA-DFT\cite{perzun}
and, for some selected clusters (e.g., $Na_8$), lead to
ground-state structures which are distinct from the LDA-DFT results.
The energy differences between close lying isomers of
an alkali cluster are often associated with
charge redistribution
and charge polarization phenomena\cite{weare}, which
are relevant,
not only to distinguish between isomers of the same cluster, but also
to understand the difference between the ground-state geometry of
isoelectronic clusters of different alkali species\cite{weare}. Since these
are
aspects in which the self-interaction effects play a role,
it is important to have an estimate of the size of these errors.
In this Letter this problem is addressed by
combining two well tested methods:
Pseudopotential Perturbation Theory\cite{eguiluz,skeleton} (PSP-PT) and
the Self-Interaction Correction (SIC)
as proposed in ref.\cite{perzun}, together with its extension
to linear response theories, developed in ref.\cite{annalen}. It
will be applied to optimize the geometries of
two small neutral sodium clusters, with
8 and 20 atoms, starting from geometries associated
with different isomers of these clusters, which have
been identified in previous LDA-DFT {\it ab-initio} calculations\cite{ursula}
(for the octamer, also in SCF-CI structural minimization\cite{vlasta2}).
It will be concluded that, for $Na_{20}$, SIC-PSP-PT provides
answers in quantitative agreement with the results of ref.\cite{ursula},
both for the bond-lengths and for the geometries of the low energy
isomers.
This seems to indicate that SIC does not play a sizeable role
in the structural
minimization of clusters with as many as 20 sodium atoms.
However, for $Na_{8}$, the results show an energy sequence for the
three lowest isomers which does not follow the
LDA-DFT {\it ab-initio} results of ref.\cite{ursula}. As a result,
it is observed that the $T_d$ structure is favoured with respect
to the $D_{4d}$ isomer, in agreement with the
SCF-CI calculations of ref.\cite{vlasta2}.
This, in turn,  seems to indicate that self-interaction effects
play a sizeable role for the smaller clusters.
\par
Pseudopotential perturbation theory has been applied to metallic
clusters in ref.\cite{skeleton}, using LDA-DFT. Being an
approximation to the more sophisticated (and more
time-consuming) {\it ab-initio} methods,
it has been shown\cite{skeleton} to provide answers in quantitative agreement
with these, in particular with {\it state-of-the-art} Car-Parrinello
structural minimization\cite{ursula}.
\par
The SIC {\it prescription}\cite{perzun}, though  lacking
a first principles justification,
has been applied with great success
in different areas of physics and chemistry, including the study
of both  ground-state and excited state properties of small
metallic clusters\cite{annalen}.
At present, there is no first-principles workable
scheme which can remedy completely the self-interaction errors
in LDA-DFT. Therefore, we resort to SIC in order to estimate
these effects. In this sense, the present study, more than
conclusive, can be of guidance for future developments.
\par
Because the
LDA-PSP-PT results are available for these clusters\cite{skeleton},
the role of self-interaction errors can be estimated unambiguosly,
the only possible source of inaccuracy being related to the
SIC itself.
We would like to point out that SIC has been used before in this context,
but only to test its ability to predict the bond-length
in alkali dimers\cite{wanda,martins}. In such calculations,
the role of SIC in the chemical bonding of atoms in a cluster cannot be
investigated.
\par
On the foregoing, we briefly summarize the SIC-PSP-PT formulation
used, before discussing the results and writing down the conclusions.
A more detailed and extended analysis will be published elsewhere
\cite{futur1}.
\par
By replacing the nuclei and core electrons by pseudopotentials, leaving
as active quantal particles the valence electrons,
one can write, for the ionic contribution to the total potential,
\begin{equation}
v_{ex}({\vec r},{\vec R})  = \sum_{i=1}^{N} v_{ps}({\vec r}-{\vec {R}_i})
\qquad , \qquad {\vec R} = \bigl\{{\vec R}_i\bigr\} \quad .
\end{equation}
In this equation, $N$ is the number of ions (also the number of
valence electrons for the neutral clusters considered here), and ${\vec {R}}$
represents a set of given
ionic positions, which ultimately should be determined by minimizing the
total energy of the system.
$v_{ps}({\vec r}-{\vec {R}_i})$ represents the
pseudopotential at point ${\vec r}$, centered at atomic site ${\vec {R}_i}$.
We shall adopt the local pseudopotential used in ref.\cite{skeleton},
with the parametrization adequate for sodium, taken from ref.\cite{uzi}.
With respect to any chosen center of the cluster, the sum of the
pseudopotentials can be expanded as follows:
\begin{eqnarray}
v_{ex}({\vec r},{\vec R} )
  &=&  \; \; \; \; \; \, v({\vec r},{\vec R}) \; \; \; \; \; \; \; \; \; +
\; \; \; \; \; \; \; \; \; v_{2,ex}({\vec r},{\vec R})
\nonumber \\
 &=&  v_0(r,{\vec R})\, Y_{0,0}(\hat r)\nonumber \\
&&  \; \; \;  +  \; \; \;
\sum_{l=1}^{\infty} \sum_{m=-l}^l v_{l,m}(r,{\vec R})\, Y_{l,m}(\hat r) \; .
\end{eqnarray}
The first term is just the monopole, spherical part of the total
ionic contribution, which will be taken into account exactly in
SIC-LDA by solving the
set of SIC Kohn-Sham-like equations\cite{perzun,annalen} for the $N$
valence electrons moving
selfconsistently in this spherical potential\cite{note}. The remaining terms
are included perturbatively up to second order,
which leads to a contribution $\Delta E_{ps}^{(2)}$ to the
total energy reading\cite{skeleton},
\begin{equation}
\Delta E_{ps}^{(2)} = {1\over 2} \int d{\vec r} \,\delta
n_2({\vec r},{\vec R})
                  \, v_{2,ex}({\vec r},{\vec R}) \; ,
\end{equation}
where $\delta n_2({\vec r},{\vec R})$ is the screened induced
density change caused by the external potential $v_{2,ex}({\vec r},{\vec R})$.
The total energy is then given by the sum of the monopole part, obtained
via the SIC-LDA self-consistent solution, plus the term $\Delta E_{ps}^{(2)}$
given by eq.(3). As in ref.\cite{skeleton}, the
screened induced density will be computed making use of
linear response theory, which incorporates, in a
self-consistent manner, the screening of the external perturbation
by the valence electrons. Since
$\delta n_2({\vec r},{\vec R})$ is built out of SIC-LDA spherical
self-consistent solution, it is in fact computed
in a numerically exact way via Green's functions techniques (for
an elaborated discussion of the methods and techniques, the reader
is referred to refs.\cite{annalen,futur1}).
As shown in ref.\cite{annalen}, the
standard LDA-DFT linear response theory, known as Time-Dependent
LDA (TDLDA), introduces a spurious self-polarization effect, which leads to
an erroneous overscreening of the external field.
The extension of SIC to the linear response theory has been formulated
in ref.\cite{annalen}, and it is this SIC response formulation
which is used in computing
$\delta n_2({\vec r},{\vec R})$\cite{annalen}.
\par
Instead of performing a fully-relaxed structural optimization,
we started from the given LDA-DFT low-energy isomeric structures
found in ref.\cite{ursula}, and then minimize the total energy using
SIC-PSP-PT by scaling uniformly the radial distance of the
constituent atoms from the center of mass of the cluster. This
strategy (similar to those adopted in refs.\cite{skeleton,vlasta2}),
being incomplete, will prove sufficient for the present purpose.
In table 1 are given the results obtained with this methodology for
$Na_{20}$. Two low energy isomers have been identified in
ref.\cite{ursula}, labelled $A$
and $B$, respectively. The $A$ structure corresponds to the lowest isomer,
with
an extra binding with respect to the $B$ structure of $0.136$ eV\cite{ursula}.
In the first row the PSP-PT results obtained in
ref.\cite{skeleton} are shown for these 2 isomers, within LDA-DFT,
using the minimization
strategy adopted here. In the second row, the SIC-PSP-PT
results of the present calculation are tabulated.
All the ingredients of the 2 calculations are the same, except for SIC.
Furthermore, and for each rowm the energy differences are always referring to
the structure which has minimum energy. Furthermore,
the numbers in parenthesis indicate the value of the scaling factor
at which a given structure minimizes its energy. One can observe that
SIC leads to a (small) reduction of the bond length in
comparison with PSP-PT, in
agreement with the previous results for dimers\cite{wanda,martins}.
More important, however, is the fact that SIC-PSP-PT keeps the same
energy ordering for the 2 isomers, therefore
evidencing {\it no} measurable effect, apart from the
expected difference in the restoring
force coefficient for the breathing mode\cite{futur1}.
\par
The situation is quite different
for $Na_8$. The results are shown in
table 2, where we keep the same notation as in table 1.
The 3 structures considered are identified in fig.1.
Whereas SIC-PSP-PT leads to structures in which the volume
of delocalization of the valence electrons, controlled by the
scaling factor, is in excellent agreement
with the results of ref.\cite{ursula}, the energy ordering
of the isomers is different in SIC-PSP-PT, when
compared with the predictions of refs.\cite{skeleton,ursula}.
This, in view of the ingredients of the calculations used,
turns out to be a pure self-interaction effect. Interestingly,
the results obtained with the present method show a
favouring of geometries which correlates well with the results
of ref.\cite{vlasta2}, which makes use of a trully self-interaction free
theory. This, in turn, reinforces the results of the
present calculation, which shows that self-interaction effects should
not be overlooked in the important issue of predicting
the lowest isomeric structures of small alkali clusters. In this
context, it is worth mentioning the recent
measurements of the photoabsorption cross-section of $Na_9^+$
clusters\cite{hh} in a
cold beam, which reveals a
multi-peak structure which can be reproduceded\cite{futur2} by
computing the absorption from the
ground-state geometry found with
the self-interaction free
SCF-CI\cite{vlasta1}, after
an appropriate scaling of the bond-lengths.
\par
Finally, we would like to comment on the average bond-length
obtained using the present calculation. The present
results lead to bond-lengths in quantitative agreement
with the findings in refs.\cite{skeleton,ursula},
as opposed to the results of SCF-CI\cite{vlasta1}, which lead to
an average bond-length in small alkali clusters which is
larger, on the average, that what is known in the bulk
limit\cite{ursula}. As the cluster size decreases, one expects
a contraction of the average bond-length, due to the corresponding
increase of the surface to volume ratio. This qualitative
reasoning, as well as the comparative performance of
LDA-DFT methods and SCF-CI methods for the sodium dimer (cf. ref.\cite{ursula})
lend support to the results of the present calculation.
In any case, the reason behind this difference
cannot be attributed to a self-interaction problem but, instead,
to the different treatment of correlations between the active
electrons.
\par
In summary, in what concerns the geometry optimization of small
alkali metal clusters, one has shown that self-interaction effects
cannot be overlooked, and may be crucial for an accurate
assignment and sorting of the low energy isomers. This feature decreases with
increasing cluster size, and seems to be unimportant already for $Na_{20}$.
As is well known\cite{annalen}, self-interaction effects still manifest
themselves in other properties of these clusters, such as the energy of the
one-electron orbitals, modifying also the topology of the
potential energy surface, which is reflected in a different
vibrational spectrum\cite{futur1}.
\par
\vspace{1.0cm}
\par
Financial support by JNICT through the project
PBICT/FIS/1635/93 is gratefully acknowledged.
The authors are
indebted
to Ursula R{\"o}thlisberger and Wanda Andreoni for providing us the
results of their Car-Parrinello total energy minimization.

\begin{table}
\caption[]{
Results for $Na_{20}$, for the two lowest isomers found in ref.\cite{ursula},
obtained using standard PSP-PT\cite{skeleton}, and the SIC-PSP-PT developed
in this work. For each of the 2 geometries selected, two sets of values are
given, associated to the result of a total energy minimization as a function
of the scaling parameter. The value of the scaling parameter at minimum is
given in parenthesis, whereas the value of the total energy constitutes the
other result tabulated. The values quoted for the energies (eV) are referred
to the state which, for each method, corresponds to the absolute minimum.
}
\begin{center}
\begin{tabular}{l c c}
$       Na_{20}  $&$       A       $&$        B       $\\ \tableline
  {\bf PSP-PT}    &$ 0.000 (1.02) $&$  0.111 (1.01) $ \\
{\bf SIC-PSP-PT}  &$ 0.000 (1.00) $&$  0.237 (1.00) $ \\
\end{tabular}
\end{center}
\label{tab1}
\end{table}

\begin{table}
\caption[]{
Results for $Na_{8}$, for the three lowest isomers found in
refs.\cite{ursula,vlasta2}, illustrated in fig.1. The notation used here is
the same as in table 1.
}
\begin{center}
\begin{tabular}{l c c c}
$       Na_8     $&$    D_{2d}     $&$      T_d     $&$     D_{4d}    $\\
\tableline
  {\bf PSP-PT}    &$ 0.000 (1.02) $&$  0.159 (1.02) $&$  0.111 (1.01) $ \\
{\bf SIC-PSP-PT}  &$ 0.000 (1.00) $&$  0.104 (1.00) $&$  0.159 (1.00) $ \\
\end{tabular}
\end{center}
\label{tab2}
\end{table}

\begin{figure}
\ \vspace{2cm}\\
Figure available from: \verb|ekardt@fhi-berlin.mpg.de|
\ \vspace{2cm}\\
\caption[]{
The 3 different geometries obtained in ref.\cite{ursula} for
$Na_{8}$, corresponding to the ground-state and two lowest
isomers (energy is increasing from left to right), and used as
starting geometries which are minimized as a function of a
dimensionless scaling parameter, which
uniformly and simultaneously changes the radial distances
of all the atoms with respect
to the center of mass of the cluster. The structure on the left
displays $D_{2d}$ symmetry, the one in the middle $D_{4d}$
symmetry, and the one on the right $T_d$ symmetry (see main
text for details).
}
\label{fig1}
\end{figure}


\begin{references}
\bibitem{note}      Although in refs.\cite{skeleton,futur1} it is shown
                    that PSP-PT and SIC-PSP-PT work well also for open-shell
                    systems, the choice of closed-shell systems adopted in
                    this work (closed-shell in a spherical jellium sense),
                    with 8 and 20 valence electrons, maximizes the
                    perturbative role played by $v_{2,ex}({\vec r},{\vec R})$.
\end{references}
\end{document}